\documentclass[final,5p,times,twocolumn]{elsarticle}

\usepackage{amssymb}
\usepackage{amsthm}
\usepackage{amsmath}
\usepackage{hyperref}
\usepackage{picture}
\usepackage{color}
\usepackage{nth}
\usepackage{gensymb}
\usepackage{lineno} 

\journal{Nuclear Instruments and Methods in Physics Research A }

\begin{document}

\begin{frontmatter}



\title{A complementary compact laser based neutron source}
 
\author[a]{A. Cianchi}\ead{alessandro.cianchi@uniroma2.it}
\author[a]{C. Andreani}
\author[b]{R. Bedogni}
\author[c]{G. Festa}
\author[b]{O. Sans-Planell}
\author[a]{R. Senesi}

\address[a]{University of Rome Tor Vergata and INFN-Roma Tor Vergata, Via della Ricerca Scientifica 1, 00133 Rome, Italy}
\address[b]{INFN-LNF, Via Enrico Fermi 40, 00044 Frascati, Italy}
\address[c]{Centro Fermi, Piazza del Viminale, 1, 00184 Roma, Italy}


\begin{abstract}
Several experiments of neutron generation using high intensity laser sources, with a power exceeding 10$^{19}$W/cm$^2$ via TNSA (Target Normal Sheath Acceleration) or other similar methods, have been performed in the past years in different laboratories. However, so far there is no one running neutron source based on such a technology. In the framework of the Conceptual Report Design of a new accelerator in the Eupraxia project \cite{eupraxia} we are studying the possibility to have a laser-based neutron source, not only by TNSA but also from self-injection schemes. We focus our attention on the applications in cultural heritage studies as well also on the complementary role that such a source can have in the framework of large facilities devoted to radiation production.

\end{abstract}

\begin{keyword} Neutron source, Cultural heritage, high power laser


\end{keyword}

\end{frontmatter}
\section{Introduction}
\label{sec:introduction}
The need for accelerating gradient order of magnitudes larger than existing ones drives the research in particle accelerators to plasma-based accelerators \cite{faure}, \cite{leemans2014multi}, \cite{muggli2009review}. 
These machines can sustain field greater than tens of GV/m paving the way to the realization of tabletop accelerators, that can successfully used for several applications, like medicine, material inspection, cultural heritage studies, radiation sources, basic research. 

Two main schemes are under consideration: Laser Wakefield Plasma Acceleration (LWPA) \cite{leemans2006gev}, without the need of any RF (Radio Frequency) conventional accelerator, or Plasma Wakefield Accelerators (PWFA) \cite{litos2014high} - \cite{clayton1996generation} where both a high intensity laser and high brightness electron beam are foreseen. 
In both schemes, a PW class laser is used.
Likely, these lasers will not be considered uniquely to drive the main accelerator but they will also be devoted to other several activities. 
The interaction of such a laser with the matter produces a large number of electrons, ions, positrons, protons via several different mechanisms depending on the laser intensity and the target material compositions and dimensions \cite{ledingham2010laser}. 
While a lot of effort is dedicated nowadays to improve the quality of these charged or neutral beams of particles to use them, so far there is not yet a user facility using them.
There are many studies concerning neutron production via laser-matter interaction (for instance among the others \cite{roth2013bright}, \cite{kar2016beamed}). 
The purpose of this paper is to consider if the laser interested for these future accelerators, and for the EUPRAXIA case for instance, can be used to drive a user-oriented compact neutron source. 
With the words user-oriented, we mean a facility where neutron flux and spectrum are enough to be competitive with existing facilities \cite{anderson2016research}, or in any case with neutron portable devices. 
One of the main application that we can foresee for such kind of source is non destructive inspections for industry, research and cultural heritage. 

Usually many techniques are used to study the objects in this field, like THz, IR and X-ray radiation. 
All of these sources are foreseen, sooner or later, in Eupraxia project. 
Adding also, within the same instrumentation, a neutron source could be very interesting, giving the possibility of having on the same site all these techniques together, in views of an integrated suite of light and particle beams for materials and cultural heritage sciences. 

This would provide the complementary high penetration, isotope selectivity, and non destructive character of neutron based techniques within the suite of light and particle probes available within the proposed project.
This neutron source could be fast (multi MeV) or, through the application of a dedicated compact moderator, thermal. 
The following analysis refers, as an exemplificative target, to the thermal neutron radiography. 
This application requires, in addition to a neutron source, a moderator. 
Further equipment as collimators and imaging systems are also required. 
In addition, a Prompt Gamma Activation Analysis (PGAA) system may be easily incorporated in the design, for expanded capability towards isotopic and elemental sensitivity.

\section{Methodological analysis}
A very compact neutron moderator is required to produce a small sized radiography facility. 
While a custom design for such kind of source should be considered, we use here as an example the scheme HOTNES (Homogeneous Thermal NEutron Source) \cite{bedogni2017experimental}, implemented already in a thermal neutron irradiation facility with extended and very uniform irradiation area (HOTNES at ENEA-INFN, Frascati).
Starting from a cm-sized fast neutron source ($^{241}$Am-B), this new type of moderator produces a highly thermalized and very uniform neutron field across a large irradiation area (30 cm in diameter). 
The moderating efficiency (thermal fluence per primary neutron) slightly depends on the fast neutron spectrum and is in the order of  2 10$^{4}$ cm$^{-2}$. 
For the purposes of the current project idea, fast neutrons from the laser compact source would be fed into the moderator, instead of the radionuclide neutron source.

\section{Possible sources}

Several possibilities can be explored, relying on primary electron, protons, or ions. These particles can be produced with the conventional RF linac or with a laser based machine.

\subsection{RF linac based source}

We consider the electron beam of 1 GeV energy, with bunch charge 100 pC, a repetition rate of 10 Hz and an average current of 1 nA. 
We use the production of neutron via bremsstrahlung, using a tungsten target of 5x5x9 cm$^{3}$, being 9 cm the thickness that maximize the yield. 
We can obtain about 0.4 neutrons for primary electron. 
With the Eupraxia design values it reflects in about 2.5 10$^{9}$ neutrons/s, which is similar to the yield of high-performance industrial neutron generators. 

A HOTNES-like moderator \cite{bedogni2017experimental} would convert them into thermal neutrons with efficiency 2.3 10$^{-4}$ cm$^{-2}$, producing a final non-collimated thermal fluence of about 6 10$^{5}$. 
Collimating devices would then reduce it of about one order of magnitude. 
In terms of overall efficiency, the non collimated thermal fluence per primary electron on the W target would be in the order of 9 10$^{-5}$ cm$^{-2}$. 

\subsection{Protons/ions from laser acceleration}

When a high intensity laser (10$^{19}$ W/cm$^{2}$ or better) is focused on a solid target several effects can be triggered, like for instance Target Normal Sheath Acceleration (TNSA) \cite{hatchett2000electron}, Radiation Pressure Acceleration (RPA) \cite{esirkepov2004highly}, collisionless shock acceleration \cite{haberberger2012collisionless} and Break Out Afterburner
(BOA) (\cite{henig2009enhanced}-\cite{yin2006gev}), depending on target material, thickness and surface contaminations for instance. 

Let us focus on TNSA. 
Fast electrons are accelerated through the material by the laser. 
These electrons penetrate the target ionizing other particles and escaping from the other side. 
In this moment, they build up a very strong electric field, in the order of TV/m. 
This field extracts protons and ions from the rear surface, producing an intense beam of particles.

\begin{figure}[htb]
  \centering
  \includegraphics*[width=80mm]{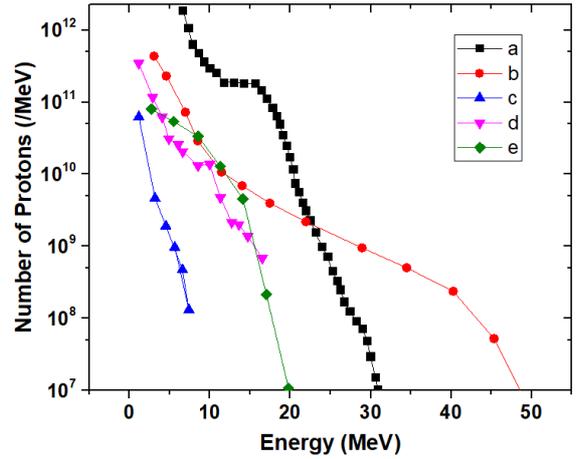}
  \caption{Comparison of proton spectrum emerging in different experiments. For the symbol 
caption see the Table 1}
  \label{fig1}
\end{figure}

While there are scaling laws of the process, being exhaustively reported in \cite{fuchs2006laser}-\cite{zeil2010scaling}, it is very difficult to define the energy spectrum, the flux intensity, and the particles geometrical distribution in a general case, being the emission strongly linked to target material, surface contamination, laser energy and intensity, laser contrast. 
The increasing of the laser energy increases both protons number and average energy. 
The great varieties of results can be appreciated in Fig.\ref{fig1}.

\begin{table}[!htbp]
\begin{center}
\begin{tabular}{|c|c|c|c|}
\hline
Label & Intensity (W/cm$^{2}$) & Energy(J) & Reference \\
\hline
a) & 2.0e20 & 200 & \cite{yang2004neutron} \\
\hline
b) & 1.5e20 & 80 & \cite{gaillard2011increased} \\
\hline
c) & 1.0e20 & 3 & \cite{kar2016guided} \\
\hline
d) & 1.0e20 & 42 & \cite{clark2002measurements} \\
\hline
e) & 1.0e21 & 10 & \cite{green2014high} \\

\hline
\end{tabular}
\caption{\label{tab:III}. Symbol caption for spectrum in Fig.\ref{fig1}.}
\end{center}
\end{table}

We can see that the laser energies are quite different, while the intensities are much more similar.
However, this dependence is not followed very strict, mainly due to some particular experimental arrangement of the target, in order to increase and to guide better the protons, as wells as the use of different kind of targets triggering mechanisms different from TNSA for instance.

What is important is the number of the proton in an energy range below 10- 20 MeV, because even laser with energy around or below 40 J can already produce a protons number in the order of 10$^{11}$ in such energy range. 
We will see in the following that the neutron yield does not change so dramatically when we consider protons of energy in this energy range with respect to higher energy values.

\begin{figure}[htb]
  \centering
  \includegraphics*[width=80mm]{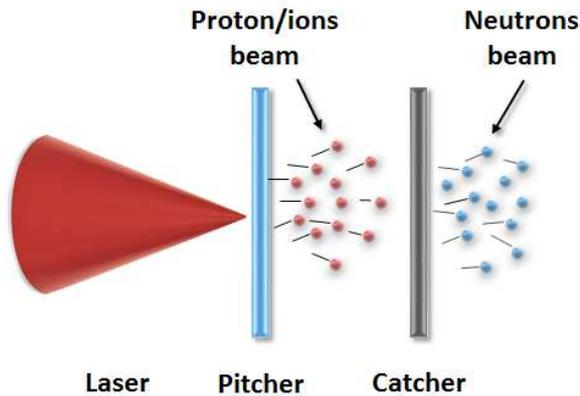}
  \caption{Sketch of the Pitcher-Catcher system}
  \label{fig2}
\end{figure}

Once that the primary beam is produced, the protons/ions hit a materials like for instance LiF or Be, in order to produce a neutron flux. 
This scheme is usually called pitcher-catcher scheme, as shown in Fig.\ref{fig2}.

We focus on a typical target of Lithium fluoride with the reaction $^{7}$Li + p $\rightarrow$ $^{7}$Be + n - 1.644 MeV. 
Different yields are found in literature for the same reaction. 
For this simulation and for the following, the neutron production yield was generated using a custom Labview$^{TM}$ based software based on a continuous projectile slowing down in the target. 
The stopping power data were generated from PSTAR (NIST) \cite{pstarurl} for protons and SRIM 2011 \cite{srim} for deuterons. 
Cross sections data were taken from ENDFB VII \cite{chadwicknucl} and EXFOR \cite{exfor} (protons) or TENDL2009 (deuterons) \cite{koning2009tendl}. 
The differences in the yield in the range of the tens-few tens of MeV are only about a factor 2-4, much less than an order of magnitude, making appealing also proton of lower energy that can be produced by smaller laser with higher repetition rate. 
The thickness of the target was also optimized in order to maximize the neutron flux. 

With 0.2 cm of LiF target the typical moderation yields in thermal neutron flux per primary fast neutron is 1.55 10$^{-4}$ n/p per proton of 5 MeV. 

We also considered the case of a Beryllium target with a deuteron beam because usually the targets are hydrogenated on the surface to increase the deuterons number. 
Considering 7 MeV deuteron we can have a moderation yields in thermal neutron flux per primary fast neutron of about 1.24 10$^{-4}$ n/d.
We are neglecting here the slightly different moderator efficiency of 7 MeV deuteron. Also in this case a custom and more refined designed should be implemented. However, we are considering just orders of magnitude to have a better understanding of the several possibilities offered by the actual technology.

\subsection{Electrons from laser acceleration}

There is another possibility, not yet fully explored, to produce neutron by a laser source: using electrons from self-injection \cite{pomerantz}. 
A TW class laser is focused in a tight spot (few $\mu$m) on a supersonics gas jet. 
The laser ionizes the plasma and the ponderomotive force remove the plasma electron generating an intense electric field. 
Inside this bubble structure, the electrons are self- injected from the rear of the bubble experiencing a strong accelerating field.

If we consider having electron of about 250 MeV on tungsten, we can have about 0.08 neutron per primary particle, and about 0.4 at 1 GeV electron energy. 

Regarding the charge there are scaling laws, as reported in \cite{lu2007generating}, being the total number roughly proportional to the square of the laser power. 
With 1 PW laser about 1.2 nC of electrons can be produced. 
However, gas mixture can increase the number as recently proved in \cite{schramm2017first}, where they obtained with a 200 TW laser about 0.5 nC at 250 MeV, with 20 $\%$ energy spread, using several mixtures of helium and nitrogen. 

Such kind of laser, similar to FLAME already existing in Frascati, can have a repetition rate of 10 Hz. 
Being it an experimental result we can use with baseline for our evaluation, obtaining about 2.5 10$^{9}$ n/s before the moderation. 
The increase of the energy is also a possibility. 
In \cite{leemans2014multi} a beam of 4.2 GeV has been produced, even at low charge. 
The main difficulty in increasing the bunch energy is the dephasing between laser and electrons. 
It cannot be really addressed in a gas jet and more complex structures, like a capillary with modulated density profile, are needed. 

\begin{table*}[t]
\begin{center}
\begin{tabular}{|c|c|c|c|c|c|c|}
\hline
Source & Primary & Energy (MeV) & Y(n/prim) & m(moderation efficiency) & Yxm & Neutrons/s/cm$^{2}$\\
\hline
RF & Electrons & 1000 & 4.0e-1 & 2.3e-4 & 9.3e-5 & 5.8e5 \\
\hline
Laser & Electrons & 250 & 8.0e-2 & 2.0e-4 & 1.6e-5 & 5.0e5 \\
\hline
Laser & Electrons & 1000 & 4.0e-1 & 2-0e-4 & 8.0e-5 & 3.0e6 \\
\hline
Laser & Protons & 5 & 8.7e-4 & 2.2e-4 & 1.9e-7 & 2.0e5 \\
\hline
Deuterons & Protons & 7 & 7.6e-4 & 1.2e-4 &  9.4e-8 & 9.4e4 \\
\hline
\end{tabular}
\caption{\label{tab:I}Uncollimated thermal neutron fluence rate expected from different fast neutron sources. For Proton and Deuterons we assume 10$^{11}$ particles per shot at 10 Hz, for laser electron 0.5 nC at 10 Hz for the 250 MeV case, while 1.2 nC at 5 Hz for 1 GeV case.}
\end{center}
\end{table*}

In Table 2 we collected the results of the simulation from different possible sources. For protons and deuterons, we estimated 10$^{11}$ particles per shot and at 10 Hz, a result that today it is the borderline of what it is possible to obtain. 
For electron, we assumed the case of \cite{schramm2017first} being already obtained. 
Laser and RF source (with the EUPRAXIA parameters) are giving roughly the same amount of neutrons. However, an improvement of the laser system to the PW scale or to the self-injection mechanics can really increase this number of a consistent factor. There are also two considerations about this number. 
First, the number from RF conventional source is the maximum achievable, today and likely in the future if the accelerating charge or the repetition rate will not increase, while the laser-based techniques are quite young and we expect to have an increase in these numbers in the following years. 
Second: even today, all of these solutions can drive easily a compact neutron source.

\section{Conclusion}

With modern neutron imaging systems, neutron radiography can be performed with a parallel beam of low energy neutrons with fluence rate 10$^{4}$ - 10$^{6}$ cm$^{-2}$s$^{-1}$.
PGAA (Prompt Gamma Activation Analysis) are less demanding considering that even conventional portable sources are used for this end, giving flux on the sample in the order of 10$^{3}$ n/cm$^{2}$/s. 
These kind of numbers are in the same order of several CANS \cite{anderson2016research}.  
Investigations related to cultural heritage may represent a strong asset of the potential as a user-oriented facility. 
These can benefit from a large and easily accessible inventory of cultural heritage artefacts in the regional area where EUPRAXIA will be implemented, fostering the access to such an infrastructure of users from museums, cultural heritage research centers, conservation and restoration centers. 
It is expected that these activities will in turn attract users from the industrial and research based community.

Both of these techniques are widely used in cultural heritage studies and can successfully implemented in EUPRAXIA using the existing infrastructure. 
Laser based source have the great advantage to be very compact, to do not require the beam of the main machine, and in prospective they will allow to deliver brighter flux of neutrons. 

However, electrons from self-injection are a great candidate to drive this research, requiring a modest laser energy, a simple setup, and they can produce enough neutrons for cultural heritage applications, that can benefit also from the presence of the radiation sources in the entire spectrum in the EUPRAXIA facility. 

\section{ACKNOWLEDGEMENTS}

This work was supported by the European Union‘s Horizon 2020 research and innovation programme under grant agreement No. 653782.


\bibliographystyle{elsarticle-num}



\end{document}